\documentclass{llncs}
\pdfoutput=1
\usepackage{graphicx}
\usepackage{grffile}
\usepackage{longtable}
\usepackage{wrapfig}
\usepackage{rotating}
\usepackage[normalem]{ulem}
\usepackage{amsmath}
\usepackage{textcomp}
\usepackage{amssymb}
\usepackage{hyperref}
\usepackage[binary-units]{siunitx}
\usepackage{tikz}
\usepackage{tikz-cd}
\usepackage[lined,boxed,linesnumbered]{algorithm2e}
\usepackage{mathtools}

\subtitle{A generalised physics of cognition}
\usepackage[style=lncs, sortlocale=en_AU, doi=true, url=false, hyperref=true, backref = true, backend=biber, natbib=true] {biblatex}
\usepackage{orcidlink}
\institute{School of Mathematics, Queensland University of Technology, Australia \and Respiratory Medicine Unit, Nuffield Department of Medicine, University of Oxford, UK \and Centre for Clinical Research, University of Queensland, Australia}
\usepackage{quiver}
\usepackage{subcaption}
\usetikzlibrary{positioning,fit,calc, patterns, cd}
\newenvironment{conditions}  {\par\vspace{\abovedisplayskip}\noindent\begin{tabular}{>{$}l<{$} @{${}={}$} l}}  {\end{tabular}\par\vspace{\belowdisplayskip}}
\author{Sophie Alyx Taylor\inst{1}\orcidlink{0000-0002-2418-5068} \and Son Cao Tran\inst{1}\orcidlink{0000-0002-8034-107X} \and Dan V. Nicolau Jr.\inst{1,2,3}\orcidlink{0000-0002-7758-226X}}
\date{\today}
\title{Taking Cognition Seriously}
\hypersetup{
 pdfauthor={Sophie Alyx Taylor},
 pdftitle={Taking Cognition Seriously},
 pdfkeywords={},
 pdfsubject={},
 pdfcreator={Emacs 28.0.50 (Org mode 9.4.6)}, 
 pdflang={English}}
\begin{document}

\maketitle
\begin{abstract}
The study of complex systems through the lens of category theory consistently proves to be a powerful
approach. We propose that cognition deserves the same category-theoretic treatment. We show that by considering a highly-compact cognitive system, there are fundamental physical trade-offs resulting in a utility problem.
We then examine how to do this systematically, and propose some requirements for "cognitive categories", before investigating  the phenomenona of topological defects in gauge fields over conceptual spaces.
\end{abstract}

\section{Introduction}
\label{sec:orga0f6aa7}
The study of thought has occupied academia for millennia, from philosophy, to medicine, and everything in
between. We demonstrate a promising approach to take advantage of the powerful tools developed by other
disciplines in our study of cognition, by emphasising the role that category theory -- a foundation of
mathematics -- ought to play.

We first motivate our approach by considering the effects of general relativity on a highly compact cognitive
agent, and noting the structural similarities between metric field theories in physics, and the dynamics of
mental content in conceptual spaces. We feel that these similarities practically beg for the study of
field theories in \emph{conceptual spaces themselves}, and consider the topological defects that arise; interpreting
them, as is done in physics, as \emph{particles of thought}.

\section{Background}
\label{sec:orgfb2d90c}
We first recall some required background which we build upon.
\subsection{Cognitive architecture and conceptual spaces}
\label{sec:org8493cf5}
To model a dynamical system, we require specification of two things: How the data is to be represented, and
the system that operates on it. In our approach, we use conceptual spaces and cognitive architectures, respectively.
\subsubsection{Conceptual spaces}
\label{sec:orge8a0056}
\emph{Conceptual spaces}\cite{gardenforsConceptualSpacesGeometry2004,zenkerApplicationsConceptualSpaces2015}
provide a convenient middle-ground approach to knowledge representation, situated between the symbolic and
associationist approaches in terms of abstraction and sparseness. A conceptual space is a tensor product of
conceptual \emph{domains}, which can be arbitrary topological spaces equipped with certain measures. They provide a
low-dimensional alternative to vector space embeddings, which often have no clear interpretation of their
topological properties.\cite{boltInteractingConceptualSpaces2017} Conceptual spaces represent concepts as
regions in a topological space. Because (meta-)properties of the conceptual spaces depend on context, this
suggests that a proper description involves the notion of a (pre)sheaf. Conveniently, they also provide a
well-motivated, natural construction for prior probabilities based on the specific symmetries of the
domain.\cite{decockGeometricPrincipleIndifference2016}
\subsubsection{Cognitive architecture}
\label{sec:org9bf4be5}
Cognitive architectures are an system-level integration of many different aspects of ``intelligence''. It can be
useful to think of cognitive architectures as the cognitive analogue of a circuit diagram; specifying the
hardware/operating system (in the case of robotic agents) or the wetware (in the case of biological agents)
that the processes deliberate thought, intentions and reasoned behaviour occur. They generally take a
``mechanism, not policy'' approach: just like a mathematician has the same brain structure as a carpenter, a
given cognitive architecture can produce wide range of agents, depending on what has been learned.

Cognitive architecture has been studied in the context of Artificial Intelligence for at least fifty years,
with several hundred architectures described in the literature. \Textcite{lairdSoarCognitiveArchitecture2012}
describes the SOAR cognitive architecture, one of the most well-established examples in the field. Several
surveys have been presented, such as \textcite{goertzelBriefSurveyCognitive2014,kotserubaReview40Years2016}.
\Textcite{huntsbergerCognitiveArchitectureMixed2011} describes one particular architecture designed for
human-machine interaction in space exploration.

\subsubsection{Generalised Problem Space Conceptual Model}
\label{sec:orgfba8948}
\begin{figure}
\centering
\begin{tikzpicture}
[block/.style={draw, align=center, node font=\tiny},
memblock/.style = {block},
learnable/.style = {rounded corners},
ltmemblock/.style = {memblock, draw=blue!50},
data/.style = {blue},
control/.style = {orange},
line/.style={-latex}]
\node (procmem) [ltmemblock] {Procedural\\memory};
\node (semmem) [ltmemblock, right=2mm of procmem] {Semantic\\memory};
\node (epimem) [ltmemblock, above=2mm of procmem] {Episodic\\memory};
\node (emomem) [ltmemblock,right =2mm of epimem] {Emotional\\memory};
\node (longtermmem) [draw, fit=(procmem)(semmem)(epimem)(emomem), label={Long-term memory}] {};

\node (workmem) [scale=3, memblock, below= of longtermmem] {Working\\memory};
\node (ruleengine) [block,learnable, below=2mm of longtermmem.south west] {Rule\\engine};

\node (learning) [block,learnable, left= of workmem] {Learning\\mechanisms};

\node (emostate) [block,learnable, right=of workmem] {Emotional\\state};
\node (ooda) [scale=0.75,block,learnable, right=of ruleengine] {Operator\\selection};
\node (memoryagents) [block,learnable, right=of longtermmem] {Unconscious\\processing};
\node (input) [block, below=of workmem] {Input};
\node (output) [block, right=of input] {Output};
\node (others) [block, left=of input] {Other agents};
\node (env) [draw=green!50, fit=(input)(output)(others), label=below:Environment] {};

\draw [<->, data, bend right] (workmem) to (ruleengine);
\draw [<->, control, out=90, in=270]  (ruleengine.north) to (procmem.south);
\draw [<->,control] (workmem) -- (ooda);
\draw [->, data] (workmem) -- (learning);
\draw [->, data, out=90, in=180] (learning) to (longtermmem.west);
\draw [->,data] (workmem) to (emostate);
\draw [<->,control,out=90,in=270] (emostate.north) to (memoryagents.south);
\draw [<->,data] (memoryagents) to (longtermmem);
\draw[->,control, bend right] (emostate) to (ooda);
\draw [<->,data, bend right] (emostate) to (longtermmem);
\draw[<->,data, bend left] (workmem.north) to (longtermmem.south);

\draw[->,data] (input) to (workmem);
\draw[->,data, bend left] (workmem) to (output);
\draw[->,data,out=0,in=270] (env.east) to (emostate.south);
\end{tikzpicture}
\caption{\label{fig:gpscm}A slight generalisation of the Problem Space Conceptual Model. Control lines are drawn in orange, while data flow is depicted in blue. Rounded corners on a black indicate that it while it has fixed functionality, it exposes control knobs to working memory, just like Direct Memory Access in a computer.}
\end{figure}
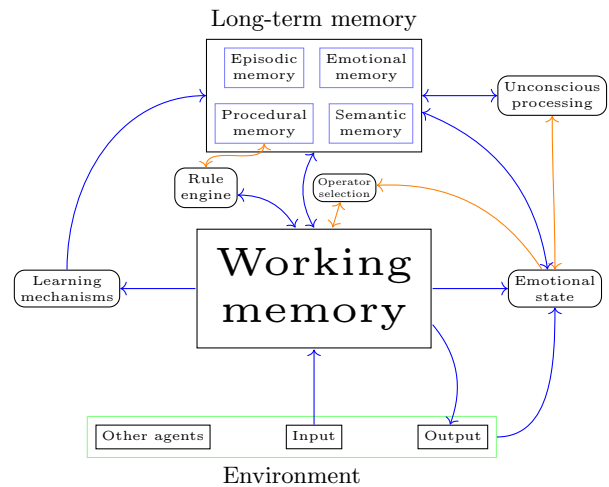

As a concrete example of a cognitive architecture, we show a modified version of the Problem Space Conceptual
Model\cite{lairdSoarCognitiveArchitecture2012} in \autoref{fig:gpscm}. It is reminiscent of a multi-head,
multi-tape Turing machine; but with transition rules stored in memory, too. The core idea of the PSCM is that
operators are selected on the basis of the contents of working memory, which is essentially short-term memory.
Production rules elaborate the contents of working memory until quiescence; operators are essentially in-band
signals about which production rules should be fired. If progress can't be made, an \emph{impasse} is generated; from
here, we can enter a new \emph{problem space}. This is essentially a formalisation of a ``point of view'' to solve a
problem.
\subsubsection{Realisation of cognitive architecture}
\label{sec:org4e4ac1e}
\begin{figure}
\centering
\begin{tikzpicture}[every node/.style = {node font=\tiny}]
\node (pscm) {PSCM};
\node (cs) [below=0mm of pscm] {Conceptual spaces};
\node (mem) [below=0mm of cs] {Memory};
\node (learn)[below=0mm of mem] {Learning};
\node (emo) [below=0mm of learn]{Emotion};
\node (ana)[below=0mm of emo] {Analogies};
\node (rule) [below=0mm of ana]{Rule engine};
\node (particles) [below=0mm of rule]{Mental instantons};
\node (att)[below=0mm of particles] {Attention};
\node (act) [below=0mm of att]{Activation};
\node (etc1)[below=0mm of act] {...};

\node (cogarch) [draw, label={Cognitive Architectures},fit=(pscm)(cs)(mem)(learn)(emo)(ana)(rule)(particles)(att)(act)(etc1)] {};

\node (info) [right=5cm of pscm] {Information};
\node (comp) [below=0mm of info] {Computation};
\node (fieldtheories) [below=0mm of comp] {Field theories};
\node (bh) [below=0mm of fieldtheories] {Black holes};
\node (physparticles) [below=0mm of bh] {Particles};
\node (symm) [below=0mm of physparticles] {Symmetry groups};
\node (coho) [below=0mm of symm] {Homology};
\node (typetheory) [below=0mm of coho] {Type theory};
\node (thermo) [below=0mm of typetheory] {Thermodynamics};
\node (proof) [below=0mm of thermo] {Proof theory};
\node (etc2) [below=0mm of proof] {...};

\node (physics) [draw, label={Mathematics and Physics}, fit=(info)(comp)(fieldtheories)(bh)(physparticles)(symm)(coho)(typetheory)(thermo)(proof)(etc2)] {};

\draw [->, bend left] (cogarch) to node [above] {Neurobiology} (physics);
\draw [->, bend left] (physics) to node [below] {Cognitive modeling} (cogarch);
\end{tikzpicture}
\caption{\label{fig:standardRealisation}The abstract models of cognitive architectures are typically concretely realised by way of neurobiology.}
\end{figure}
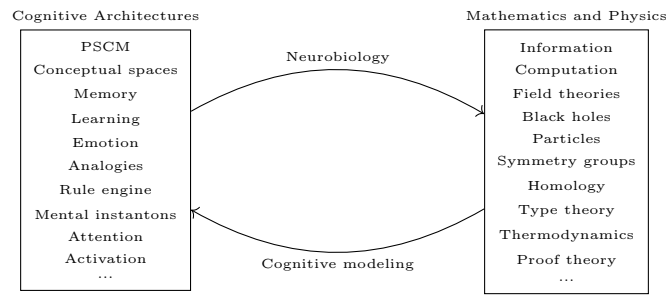

Cognitive architectures are \emph{abstract models}, which need to be \emph{concretely realised}. We almost always study
cognition through a realisation via neuroscience, which itself is realised via biology, then chemistry, then
physics. This is depicted in \autoref{fig:standardRealisation}. Alternatively, we may realise it in the form of
robotics and software simulations; like with neuroscience, these are also built upon a tower of realisations.

\subsection{Limits on computation}
\label{sec:org1fe4ba7}
Computation has its limits; most famously, the halting problem. We can find limits of computation in at least
two general classes: limits imposed by computation in-and-of-itself, and limits imposed by physics on the
embodiment of a computation.
\subsubsection{Busy-Beaver Machines}
\label{sec:org8b71535}
Limits on maximum output of a Turing machine for a given number of states, and number of transitions,
  \(BB(n)\). The Busy Beaver game is to construct an \(n\)-state Turing machine which outputs the largest
  amount of ``1''s possible, \emph{then halts}. A variation on this is to count the number of steps it takes for the
  said Busy Beaver machine to halt. If you look at this as a function in \(n\), whether counting the size of
  the output or the transitions, you have an extraordinarily fast-growing function, quickly outpacing
  Ackermann's function. You can use this to decide whether \(n\)-state Turing machines halt, simply by waiting
  BB(n) steps. If it takes longer than that, then you know it will never halt, by definition, because it would
  be the new BB(n). Of course, BB(n) itself is uncomputable in general: We have no free lunch.
\subsubsection{The Bekenstein bound}
\label{sec:org2a25103}
The Bekenstein bound is a limit on information density:
\begin{equation}
\label{eqn:bekenstein}
I \le \frac{2\pi RE}{\hbar c\ln 2} \quad\si{\bit}
\end{equation}
where \(R\) is the radius of the system, \(E\) is the energy of the system, \(\hbar\) is the reduced Planck's
constant, and \(c\) the speed of light.\cite{2005RPPh...68..897T} If one assumes the laws of thermodynamics,
then all of general relativity is derivable from \autoref{eqn:bekenstein}.\cite{1995PhRvL..75.1260J} If you
fixes the radius \(R\), then as you try to fit more and more information in the given volume, you must
eventually start to increase the energy in that volume. Eventually, the Schwarzchild radius of that energy
equals \(R\), and you have a black hole. This occurs when \autoref{eqn:bekenstein} is an equality.
\subsubsection{The Margolus-Levitin bound}
\label{sec:orge5d94e6}
The Margolus-Levitin bound is a bound on the processing rate of a system. A (classical) computation must take
at least \(\frac{h}{4E}\). To decrease the minimum time to take a computational step, you must increase the
energy involved. If this energy is contained in a finite volume, then you suffer the same problem arising from
the Bekenstein bound; eventually the Schwarzchild radius catches up to you. It's important to note, however,
that this assumes no access to quantum memory.

\subsection{Gauge fields and their symmetries}
\label{sec:org3dbd856}
\begin{figure}
\centering
\begin{tikzcd}
	E & \simeq & {\Sigma \times F} \\
	\\
	\Sigma
	\arrow["\sigma", curve={height=-6pt}, from=3-1, to=1-1]
	\arrow["fb", curve={height=-6pt}, from=1-1, to=3-1]
	\arrow["{\text{pr}_1}", from=1-3, to=3-1]
\end{tikzcd}
\caption{\label{fig:gaugeBundle}A gauge field \(F\) over a spacetime \(\Sigma\).}
\end{figure}
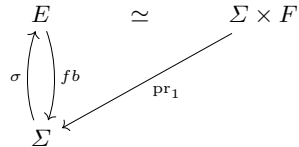

A \emph{gauge field} is an approach to formalising what we mean by a \emph{physical field}. We think of a gauge field as
data ``sitting above'' the points in a space (a \emph{principle bundle}), as shown in \autoref{fig:gaugeBundle},
combined with a transformation rule (a \emph{connection}) to ensure the data transforms sensibly with a change of
test point.\cite{baez1994gauge} Gauge fields typically have a very useful property: We can reparameterise them
to be more convenient to work with. For example, we can change the electric and magnetic potentials in an
electromagnetic field by a consistent transformation derived from arbitrary data, yet the exact same physical
effects will occur. This is called \emph{gauge fixing}.

\subsection{Topological defects}
\label{sec:org2295374}
Topological defects are an important notion in a number of scientific and engineering fields. Intuitively
speaking, they are \emph{imperfections} in what we call an \emph{ordered medium}; roughly, something which has a consistent
rule for assigning values to a point in space, like a gauge
field.\cite{merminTopologicalDefectsOrderedMedia1979} Some familiar examples include:

\begin{itemize}
\item Grain boundaries in metals, consisting of disruptions in the regular lattice placement of the constituent
atoms,
\item Singularities and vortices in vector fields, such as black holes and tornadoes, respectively, and
\item Gimbal lock is a topological defect in the Lie algebra bundle \(\mathfrak{spin}(3) \to T^3\) of rotations
over the Euler angles.
\end{itemize}

In general, defects arise when there are non-trivial homotopy groups in the principle bundle of the gauge
field; that is, when you can make a loop in the bundle space, but you can't shrink it down to a single point.

These are non-perturbative phenomena, arising due to the very \emph{nature} of these systems. These manifest
themselves as \emph{(psuedo-)particles} or higher branes, such as membranes. However, due to the risk of confusion
between the homonyms ``brain'' and ``brane'', we will call them all particles, regardless of their dimension.
\subsubsection{Dynamics}
\label{sec:orgb55d244}
What makes defects particularly interesting is that they can evolve over time. Dynamical behaviour in
the base space can lead to dynamical behaviour of defects; for example, the movement of holes in a
semiconductor, or the orbit of a small black hole around a larger one.
\subsubsection{Temperature and spontaneous symmetry breaking}
\label{sec:org3f1aaf8}
Another common phenomenon is that there is a \emph{temperature} associated with the defect, which controls both its
dynamics and its very existence. The obvious example of a temperature parameter is thermal temperature; if you
melt a bar of iron, you won't have any more crystal structure to be defective. A more challenging one is
temperature in superconductors or superfluids; raise it too high and Cooper pairing no longer works, so you
lose superconductivity and superfluidity, respectively. These drastic changes in behaviour are \emph{phase changes}. 

Consider the opposite situation: Freezing a liquid to a solid. If you lower the temperature to a point where
defects start appearing, then those defects have to actually appear \emph{somewhere}. The configuration of the
defects will be just one possible configuration out of a potentially astronomically large space of
possibilities (such as the precise atoms involved in a grain boundary), chosen at random. This phenomena is
called \emph{spontaneous symmetry breaking}, and is responsible for the masses of elementary particles, via the Higgs
mechanism.
\subsection{Category theory and its slogans}
\label{sec:orga0b40b6}
Category theory is a foundation of mathematics, as an alternative to set theories like ZFC. Instead of
focusing on membership, category theory focuses on \emph{transformation} and \emph{compositionality}. Its methodology for
studying an object is to look at how that object relates to other objects. There are many excellent
introductions to category theory aimed at different audiences (such as
\cite{fongInvitationAppliedCategory2019,riehlCategoryTheoryContext2016,lawvereConceptualMathematicsFirst2012,spivakCategoryTheorySciences2014}).
These introduce the basic notions and show the versatility of category theory in a wide range of fields.

\subsubsection{Categories and functors}
\label{sec:org9ca6a1a}
A category is a collection of objects, and a collection of morphisms between those objects. For example, the
category of sets has sets as objects, and functions as morphisms. Different classes of category allow
different constructions; for example, monoidal categories allow \emph{pairing} of objects. You can translate between
categories with \emph{functors}, which are themselves just morphisms in the category of categories.

\subsubsection{Adjunctions}
\label{sec:org4cb13fd}
A common situation in mathematics arises when you want to translate back and forth between two types of
structures via a pair of functors, but the structures aren't isomorphic. The best we can do is approximate an
isomorphism; this is called an \emph{adjunction}.\cite{riehlCategoryTheoryContext2016} 
\subsubsection{Enrichment}
\label{sec:org8821219}
\emph{Enriched categories} are a generalisation of plain categories. Whereas in standard category theory, the
morphisms between objects are described by their homset, a category enriched in a monoidal category
\(\mathcal{V}\) has hom-objects from
\(\mathbf{Ob}(\mathcal{V})\).\cite[pages 56,139]{fongInvitationAppliedCategory2019} Ordinary categories are
just categories enriched over the category of sets. \cite[page 87]{fongInvitationAppliedCategory2019}

\subsubsection{An approach to science and mathematics}
\label{sec:orge867768}
Let us take a step back for a moment, and consider \emph{how} science should be done. Bill Lawvere offers this:
\begin{quote}
When the main contradictions of a thing have been found, the scientific procedure is to summarize them in
slogans which one then constantly uses as an ideological weapon for the further development and transformation
of the thing. \cite{lawvereCategoriesSpaceQuantity1992}
\end{quote}

Let us take this approach. What are some slogans for category theory?
\begin{itemize}
\item Better to have a nice category of objects than a category of nice objects \cite{corfieldModalHomotopyType2020}
\item Dialectical contradictions are adjunctions \cite{lawvereCategoriesSpaceQuantity1992}
\item Find and characterise the principal adjunction of any problem
\end{itemize}

We shall now provide some motivation for our work.

\section{Motivation}
\label{sec:org0c7c493}
We motivate our approach by considering arbitrary cognitive systems in terms of physics. First, we consider
the consequences of a physically dense cognitive agent; second, we consider the role self-parameterisation in
conceptual spaces.
\subsection{The high-density regime of cognition and its consequences}
\label{sec:org92741ba}
Let us consider the high-density regime of cognition; that is, where the cognitive agent's radius is close to
its Schwarzchild radius. We shall assume that the agent can learn from its experiences. As the agent
experiences the world, it may learn new information,\footnote{These experiences may allow generalisation of
existing information and thus discarding of the individual cases; but over time, there will be a point where
generalisations can't be reasonably made.} which requires storage in the agent's memory. Consider this
increase in information in the context of \autoref{eqn:bekenstein}: more and more energy will be required to
store it in a given volume. This, in turn, results in an increase in the Schwarzchild radius. As we are
already close to our Schwarzchild radius, and we presumably don't want to collapse into a black hole,\footnote{It
is ill-advised to become a black hole, if only to be able to affect the external environment in a reasonable
manner (but see \cite{blackHoleModelOfComputation})} we must increase our containment radius; this results in
both the information storage \emph{and} processing gadgets spreading out. This results in the average distance
between two pieces of information in the containment volume increasing, which in turn results in increased
average propagation delay to shunt information around. Thus, the average serial processing rate will decrease
--- in other words, \emph{learning can make you slower}. This isn't even taking into account gravitational time
dilation, which will only further enhance the slowdown relative to an outside observer, as well as the energy
required to interact with the external environment.
\subsubsection{Setting}
\label{sec:orgcd5deb8}
We consider an agent comprised of an incompressible fluid which is spherically symmetric and non-rotating, and
electrically neutral, resulting in the \emph{interior Schwarzschild metric}.\cite{Schwarzschild:1916ae} In
Schwarzschild coordinates, the line element is given by

\begin{equation}
\begin{aligned}
c^2 {d \tau}^{2} = & \frac{1}{4} \left( 3 \sqrt{1-\frac {R_S}{R_g}}-\sqrt{1-\frac{r^2 R_S}{R_g^3}} \right)^2 c^2 dt^2\\
&- \left( 1-\frac{r^2 R_S}{R_g^3} \right)^{-1} dr^2 - r^2 \left(d\theta^2 + \sin^2\theta \, d\varphi^2\right),
\end{aligned}
\label{eqn:ismLineElement}
\end{equation}
where
\begin{conditions}
(t,r,\theta,\phi) & The spacetime coordinates in \((+---)\) convention\\
R_S & The Schwarzschild radius of the agent\\
R_g & The value of \(r\) at the boundary of the agent, measured from the interior.
\end{conditions}
\subsubsection{Slowdown due to propagation delay}
\label{sec:org7f8307e}
If we consider an instantaneous path through our agent -- that is, one where \(dt = 0\) -- we can see from
\autoref{eqn:ismLineElement} that when \(R_g\) is close to \(R_S\), the radial term blows up as the path
approaches the surface; indeed, we can see that the distance required to be travelled to travel to the surface
when \(R_g = R_S\) is infinite.

\subsubsection{Slowdown due to gravitational time dilation}
\label{sec:org7ed9111}
The time dilation experienced by the agent compared to an observer at infinity increases with density.
The scaling compared to the outside observer is given by
\begin{equation}
\label{eqn:timeDilation}
\sqrt{1 - \frac{R_S}{r}}
\end{equation}
valid for \(r > R_S\).
\subsubsection{Extra energy required to interact with the external environment}
\label{sec:orge39dae1}
Not only will the extreme gravitational field reduce rate of processing, it will also increase the energy
required to interact with the external environment. Consider the case of our agent trying to communicate with
a distant observer via some protocol based on the exchange of radio signals. If there is any specification of
frequency of the carrier signal, then we must account for the gravitational redshifting of our agent's
emissions. This redshifting results in longer wavelength signals; equivalently, lower frequency signals.
Because the energy of the photons comprising the signal is directly proportional to its frequency, this can be
stated in terms of \emph{lower energy signals}. Thus, in order to ensure outbound transmissions meet the frequency
specifications of the protocol, the photons must be emitted with extra energy to overcome the extreme
curvature of the gravitational field.

In the simplest case, assume the exterior Schwarzschild metric.\cite{Schwarzschild:1916uq} To signal an outside
observer at infinity with a photon of a given energy \(E\) emitted at a radius \(R_E\), the photon will have
to be emitted with an energy increased by a factor of
\begin{equation}
\label{eqn:extraEnergy}
\frac{R_E}{\sqrt{R_E(R_E- R_S)}} -1.
\end{equation}
\subsubsection{Some tasks are too complex to be solved in a given volume}
\label{sec:org2a414ee}
We have thus seen that there is a \emph{fundamental physical trade-off} between learning, and the time it takes to
solve an arbitrary task. This immediately implies a number of things:
\begin{itemize}
\item There is a trade-off between the general capability of an agent, and its average reactivity,
\item Some complex tasks with a time requirement are simply too complicated for any agent to solve the task in a
given volume of space,
\item As you increase the average capability of a group of agents, less of them will be able to fit in a given
volume.
\end{itemize}
\subsubsection{Lattice of cognitive skills}
\label{sec:orgb220704}
We can use these limits to create various partial orders:
\begin{itemize}
\item We can rank tasks by the minimum and maximum volume of agents capable of solving it,
\item We can rank agents by their capabilities with respect to physically-embodied Busy-Beaver machines,
\item We can rank space-time volumes by combinations of the above
\end{itemize}
\subsection{Self-parameterisation of behaviour}
\label{sec:orgd009107}
Let us consider our second motivational example: the self-parameterisation of behaviour. The behaviour of an
agent depends on its accumulated knowledge up until that point, in particular, its procedural knowledge. Its
knowledge, in turn, depends on its past behaviour acquiring the knowledge. This self-referential quality
appears in physics, in the form of \emph{metric field theories}. Consider, again, general relativity: The metric
tensor controls the dynamics of a system, and the distribution of that system determines the metric tensor.
The previous example considered the \emph{effects} of general relativity on a cognitive agent, but the agent didn't
particularly play an \emph{active} role in our considerations.

But, why consider only general relativity? Can we consider other metric field theories? For that matter, do we
even have to consider \emph{physical} field theories?

\section{Goal}
\label{sec:org3079afb}
Consider the position our motivational examples has put us in: Instead of considering cognition as an embodied
agent comprised of wetware or hardware, we just re-enacted the old ``assume a spherical cow in a vacuum'' joke.
We didn't bother trying to determine \emph{how} a dense sphere of incompressible fluid could embody a cognitive
agent; it turned out we didn't even need to! We just assumed that the translation from an abstract model of
cognition, to the concrete sphere of cosmic horror, and then back again to an abstract model preserved the
semantics of cognition, whatever they may be.
\subsection{Taking advantage of tools from other disciplines}
\label{sec:orga5199c7}
Let us now state what we wish to do, in order to find a solution. We wish to use the mathematical tools from
other disciplines in order to study cognition. How is this typically done?
\subsubsection{The physics of cognition, by way of biology}
\label{sec:org5105064}
As shown in \autoref{fig:standardRealisation}, we translate through a chain of ``domains'' of science before
reaching physics. Each translation introduces an extraordinary amount of complexities that exist solely due to
choosing a specific concrete realisation \emph{at each stage}. While, of course, neurology and biology are \emph{incredibly}
important subjects of study in humans -- if only for their phenomenological observations, not to mention the
pathophysiological importance -- they are complicating factors in the study of the phenomena of cognition
in-and-of-itself.
\subsubsection{We only need to find nicer realisation morphisms which preserve behaviour}
\label{sec:orgbbfc02e}
\begin{figure}
\centering
\begin{tikzpicture}[every node/.style = {node font=\tiny}]
\node (pscm) {PSCM};
\node (cs) [below=0mm of pscm] {Conceptual spaces};
\node (mem) [below=0mm of cs] {Memory};
\node (learn)[below=0mm of mem] {Learning};
\node (emo) [below=0mm of learn]{Emotion};
\node (ana)[below=0mm of emo] {Analogies};
\node (rule) [below=0mm of ana]{Rule engine};
\node (particles) [below=0mm of rule]{Mental instantons};
\node (att)[below=0mm of particles] {Attention};
\node (act) [below=0mm of att]{Activation};
\node (etc1)[below=0mm of act] {...};

\node (cogarch) [draw, label={Cognitive Architectures},fit=(pscm)(cs)(mem)(learn)(emo)(ana)(rule)(particles)(att)(act)(etc1)] {};

\node (info) [right=3cm of pscm] {Information};
\node (comp) [below=0mm of info] {Computation};
\node (fieldtheories) [below=0mm of comp] {Field theories};
\node (bh) [below=0mm of fieldtheories] {Black holes};
\node (physparticles) [below=0mm of bh] {Particles};
\node (symm) [below=0mm of physparticles] {Symmetry groups};
\node (coho) [below=0mm of symm] {Homology};
\node (typetheory) [below=0mm of coho] {Type theory};
\node (thermo) [below=0mm of typetheory] {Thermodynamics};
\node (proof) [below=0mm of thermo] {Proof theory};
\node (etc2) [below=0mm of proof] {...};

\node (physics) [draw, label={Mathematics and Physics}, fit=(info)(comp)(fieldtheories)(bh)(physparticles)(symm)(coho)(typetheory)(thermo)(proof)(etc2)] {};

\draw [->, bend left] (cogarch) to node [above] {Neurobiology} (physics);
\draw [->, orange, thick] (cogarch) to node [above] {Toy models} (physics);
\draw [->, bend left] (physics) to node [below] {Cognitive models} (cogarch);
\end{tikzpicture}
\caption{\label{fig:newRealisation}All we need to do is to preserve the behaviours, not any particular concrete realisation strategy.}
\end{figure}
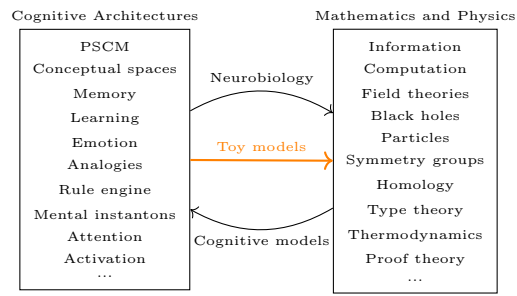

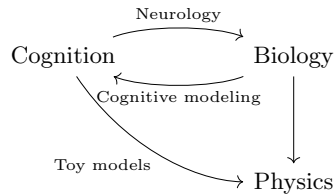
\begin{figure}
\centering
\begin{tikzcd}
	{\text{Cognition}} && {\text{Biology}}\\
	\\
	&& {\text{Physics}}
	\arrow["{\text{Neurology}}", curve={height=-12pt}, from=1-1, to=1-3]
	\arrow["{\text{Cognitive modeling}}", curve={height=-12pt}, from=1-3, to=1-1]
	\arrow[from=1-3, to=3-3]
	\arrow["{\text{Toy models}}"', curve={height=18pt}, from=1-1, to=3-3]
\end{tikzcd}
\caption{\label{fig:categoricalDiagramRealisation}Transformation expressed in a more traditional categorical diagram}
\end{figure}

Instead of considering the familiar realisation via neurobiology, the realisation transformation only needs to
preserve the structures and behaviours of interest; that is, \emph{we only need to find toy models}. To do this, we
need to find a nice class of ``cognitive categories'', and adjunctions out of them, as shown in
\autoref{fig:newRealisation}. A slightly more category-theoretic diagram is shown in
\autoref{fig:categoricalDiagramRealisation}.

\subsection{Cognitive categories}
\label{sec:org70c77c3}
If we are going to use category theory to study cognition, then we ought to specify what cognitive categories
actually are. This is an open problem, but some plausible requirements include:
\begin{itemize}
\item Subcategories of cognitive categories ought to include Turing categories, in order to capture computational
behaviour
\item There should be an opportunity for enrichment in a category of conceptual spaces.
\end{itemize}

\section{Example: Topological defects in conceptual spaces}
\label{sec:org04c9684}
Let us consider the metric field analogy a little deeper. Since many conceptual spaces of interest have a
genuine geometric structure, and we can form fibre bundles over them which allow parallel transport, then we
can at least \emph{consider} gauge fields.
\subsection{Cognitive gauge fields}
\label{sec:org79fc3bf}
There are a lot of potential options for cognitive gauge fields. Some are generic, which might apply to any
conceptual space; whereas some might only apply to certain classes of conceptual space. We assume some
mechanism for smooth interpolation in cases of noisy discrete data.

Some example plausible generic gauge fields include:
\begin{itemize}
\item The activation value of a specific memory is related to its probability of being recalled due to a query.
The higher the activation value, the more it will be recalled. These values can spread to adjacent memories.
Many ``forgetting'' mechanisms are based on forgetting memories with low activation value.
\item We can consider the emotional state of an agent when storing the memory, decomposed into a set of \emph{valuation}
and \emph{valence} affect dimensions.
\item We can also consider the subjective importance of a memory, which is separate to its activation; an example
in procedural memory is a rule saying to not completely flood a room with water if there are people in it.
It's not very \emph{likely} to be subject to recall, but it's certainly an important rule!
\end{itemize}

Some plausible non-general gauge fields might be:
\begin{itemize}
\item Trustworthiness of data gathered socially.
\item Difficulty of tasks and behaviours. This can evolve based on experience and better understanding of the
situation.
\end{itemize}
\subsection{Defects and their dynamics: Particles of thought}
\label{sec:org1ceb56c}
Some cognitive gauge fields will have non-trivial topology, resulting in topological defects. As the
underlying conceptual spaces evolve, whether contents or the topology itself, we might observe dynamical
evolution of these defects. Thus, we might (not-so-metaphorically) call these ``particles of thought''.

Leaving aside whether such a thing has a meaningful interpretation,\footnote{This almost certainly depends on the
conceptual space and gauge field involved.} we can at least ask more about the nature of such things.
\subsection{Production rules as potentials}
\label{sec:org21b446e}
Can we encode production rules as a something akin to a potential? If so, what actually generates those
potentials? 
\subsection{Transmission of influence and cognitive gauge bosons}
\label{sec:org5eb93fb}
How is the influence of a gauge field propagated? Is it wave-like, as in classical physics, or are there
'force-carrying particles', like gauge bosons in particle physics?
\subsection{Phase changes}
\label{sec:orga3dec31}
Are there phase changes? That is, is there some order parameter where the particles are only manifest in a
given range of parameter values? Does this relate to switching problem spaces?
\subsection{Cognitive event horizons}
\label{sec:org1182a73}
Are there ``cognitive event horizons'', where there are boundaries from which the effects of a particle can
never escape, not just effects on its surrounding particles or memories, but on behaviour too? If so, how do
they form? How do they evolve? Are there mechanisms to affect the underlying topological structure of a
conceptual space? Is there a ``maximum resolution'' to some conceptual spaces as a result, analogous to the
Schwarzschild radius in general relativity?
\section{Discussion}
\label{sec:orga03563d}
There is an important consideration to be made when talking about theoretical modelling: we must stress when
we are only talking about \emph{effective theories}; that is, theories which model the \emph{effects} of cognition, but do
not make any claims as to whether there is any actual \emph{causal} connection with the reality of cognition. The
ontological status of particles of thought certainly rests on the status of conceptual spaces, and gauge
fields over them. Further, assuming they \emph{are} ontologically valid, whether they actually have any causal role
requires considerable further study.
\subsection{Future research}
\label{sec:org90e2ebd}
We have a number of interesting questions prompted by our approach:
\begin{itemize}
\item How might we study the flow of information throughout an agent's lifetime? Can this be linked to ``particles
of thought''?
\item How can an agent perceive the Self? Does an agent who is aware of itself encounter, for example, the
Barber's paradox? How can it reason about things which are not true? What are the connections with
paraconsistent logic?
\item How can we more fruitfully take advantage of topological data analysis?
\item How does analogy work in our approach? Does it rely on the homology structure of the relevant conceptual
spaces having particular forms?
\item What is a good model for various learning mechanisms; both of mental content itself, and of new conceptual
spaces?
\item What group structures over different conceptual spaces can we find to yield different field theories?
\item Behaviour and cognition depends heavily on emotional state.\cite{rosalesGeneralTheoreticalFramework2019} What
is the most appropriate conceptual space to represent this, and do they have any psuedoparticles?
\end{itemize}

\printbibliography

@book{gardenforsConceptualSpacesGeometry2004,
  title =		 {Conceptual Spaces: The Geometry of Thought},
  shorttitle =	 {Conceptual Spaces},
  author =		 {Gärdenfors, Peter},
  date =		 2004,
  edition =		 {1. MIT Press paperback ed},
  publisher =	 {{MIT Press}},
  location =	 {{Cambridge, Mass.}},
  isbn =		 {978-0-262-57219-4},
  langid =		 {english},
  pagetotal =	 307,
  series =		 {A {{Bradford}} Book},
}

@collection{zenkerApplicationsConceptualSpaces2015,
  title =		 {Applications of {{Conceptual Spaces}}},
  editor =		 {Zenker, Frank and Gärdenfors, Peter},
  date =		 2015,
  publisher =	 {{Springer International Publishing}},
  location =	 {{Cham}},
  doi =			 {10.1007/978-3-319-15021-5},
  url =			 {http://link.springer.com/10.1007/978-3-319-15021-5},
  urldate =		 {2020-03-10},
  isbn =		 {978-3-319-15021-5},
  langid =		 {english},
}

@misc{boltInteractingConceptualSpaces2017,
  title =		 {Interacting {{Conceptual Spaces I}} : {{Grammatical Composition}} of {{Concepts}}},
  shorttitle =	 {Interacting {{Conceptual Spaces I}}},
  author =		 {Bolt, Joe and Coecke, Bob and Genovese, Fabrizio and Lewis, Martha and Marsden, Dan and
                  Piedeleu, Robin},
  date =		 {2017-09-29},
  urldate =		 {2020-03-10},
  archivePrefix ={arXiv},
  eprint =		 {1703.08314},
  eprinttype =	 {arxiv},
  primaryClass = {cs.LO}
}

@article{decockGeometricPrincipleIndifference2016,
  title =		 {A Geometric Principle of Indifference},
  author =		 {Decock, Lieven and Douven, Igor and Sznajder, Marta},
  date =		 {2016-12-01},
  journaltitle = {Journal of Applied Logic},
  shortjournal = {Journal of Applied Logic},
  volume =		 19,
  pages =		 {54--70},
  issn =		 {1570-8683},
  doi =			 {10/f9h7ck},
  url =			 {http://www.sciencedirect.com/science/article/pii/S1570868316300167},
  urldate =		 {2020-01-20},
  langid =		 {english},
  series =		 {{{SI}}:{{Dynamics}} of {{Knowledge}} and {{Belief}}}
}

@book{lairdSoarCognitiveArchitecture2012,
  title =		 {The {{Soar}} Cognitive Architecture},
  author =		 {Laird, JohnE.},
  date =		 2012,
  publisher =	 {{MIT Press}},
  location =	 {{Cambridge,Mass. }},
  isbn =		 {978-0-262-12296-2}
}

@incollection{goertzelBriefSurveyCognitive2014,
  title =		 {Brief {{Survey}} of {{Cognitive Architectures}}},
  booktitle =	 {Engineering {{General Intelligence}}, {{Part}} 1},
  author =		 {Goertzel, Ben and Pennachin, Cassio and Geisweiller, Nil},
  date =		 2014,
  pages =		 {101--142},
  publisher =	 {{Atlantis Press, Paris}},
  doi =			 {10.2991/978-94-6239-027-0_6},
  url =			 {https://link.springer.com/chapter/10.2991/978-94-6239-027-0_6},
  urldate =		 {2018-03-17},
  isbn =		 {978-94-6239-027-0},
  langid =		 {english},
  series =		 {Atlantis {{Thinking Machines}}}
}

@misc{kotserubaReview40Years2016,
  title =		 {A {{Review}} of 40 {{Years}} of {{Cognitive Architecture Research}}: {{Core Cognitive
                  Abilities}} and {{Practical Applications}}},
  shorttitle =	 {A {{Review}} of 40 {{Years}} of {{Cognitive Architecture Research}}},
  author =		 {Kotseruba, Iuliia and Tsotsos, John K.},
  date =		 {2016-10-26},
  url =			 {http://arxiv.org/abs/1610.08602},
  urldate =		 {2019-08-01},
  archivePrefix ={arXiv},
  eprint =		 {1610.08602},
  eprinttype =	 {arxiv},
  primaryClass = {cs.AI}
}

@inproceedings{huntsbergerCognitiveArchitectureMixed2011,
  title =		 {Cognitive Architecture for Mixed Human-Machine Team Interactions for Space Exploration},
  booktitle =	 {2011 {{Aerospace Conference}}},
  author =		 {Huntsberger, Terry},
  date =		 {2011-03},
  pages =		 {1--11},
  doi =			 {10/cq4m28}
}

@ARTICLE{2005RPPh...68..897T,
  author =		 {{Tipler}, F.~J.},
  title =		 "{The structure of the world from pure numbers}",
  journal =		 {Reports on Progress in Physics},
  year =		 2005,
  month =		 apr,
  volume =		 68,
  number =		 4,
  pages =		 {897-964},
  doi =			 {10.1088/0034-4885/68/4/R04},
  archivePrefix ={arXiv},
  eprint =		 {0704.3276},
  primaryClass = {hep-th},
  adsurl =		 {https://ui.adsabs.harvard.edu/abs/2005RPPh...68..897T}
}

@ARTICLE{1995PhRvL..75.1260J,
  author =		 {{Jacobson}, Ted},
  title =		 "{Thermodynamics of Spacetime: The Einstein Equation of State}",
  journal =		 {Physical Review Letters},
  year =		 1995,
  month =		 aug,
  volume =		 75,
  number =		 7,
  pages =		 {1260-1263},
  doi =			 {10.1103/PhysRevLett.75.1260},
  archivePrefix ={arXiv},
  eprint =		 {gr-qc/9504004},
  primaryClass = {gr-qc},
  adsurl =		 {https://ui.adsabs.harvard.edu/abs/1995PhRvL..75.1260J}
}

@book{baez1994gauge,
  title =		 {Gauge Fields, Knots And Gravity},
  author =		 {Baez, J.C. and Muniain, J.P.},
  isbn =		 9789813103245,
  series =		 {Series On Knots And Everything},
  year =		 1994,
  publisher =	 {World Scientific Publishing Company}
}

@article{merminTopologicalDefectsOrderedMedia1979,
  title =		 {The topological theory of defects in ordered media},
  author =		 {Mermin, N. D.},
  journal =		 {Rev. Mod. Phys.},
  volume =		 51,
  issue =		 3,
  pages =		 {591--648},
  numpages =	 0,
  year =		 1979,
  month =		 {Jul},
  publisher =	 {American Physical Society},
  doi =			 {10.1103/RevModPhys.51.591},
  url =			 {https://link.aps.org/doi/10.1103/RevModPhys.51.591}
}

@book{fongInvitationAppliedCategory2019,
  title =		 {An {{Invitation}} to {{Applied Category Theory}}: {{Seven Sketches}} in
                  {{Compositionality}}},
  shorttitle =	 {An {{Invitation}} to {{Applied Category Theory}}},
  author =		 {Fong, Brendan and Spivak, David I.},
  date =		 {2019-07-18},
  edition =		 1,
  publisher =	 {{Cambridge University Press}},
  doi =			 {10.1017/9781108668804},
  url =			 {https://www.cambridge.org/core/product/identifier/9781108668804/type/book},
  urldate =		 {2019-12-16},
  isbn =		 {978-1-108-66880-4},
  langid =		 {english}
}

@book{riehlCategoryTheoryContext2016,
  title =		 {Category Theory in Context},
  author =		 {Riehl, Emily},
  date =		 2016,
  publisher =	 {{Dover}},
  url =			 {http://site.ebrary.com/id/11359593},
  urldate =		 {2020-03-27},
  isbn =		 {978-0-486-82080-4},
  langid =		 {english}
}

@book{lawvereConceptualMathematicsFirst2012,
  title =		 {Conceptual Mathematics: A First Introduction to Categories},
  shorttitle =	 {Conceptual Mathematics},
  author =		 {Lawvere, Francis W and Schanuel, Stephen H},
  date =		 2012,
  publisher =	 {{Cambridge Univ. Press}},
  location =	 {{Cambridge}},
  isbn =		 {978-0-521-89485-2},
  langid =		 {english},
}

@book{spivakCategoryTheorySciences2014,
  title =		 {Category Theory for the Sciences},
  author =		 {Spivak, David I},
  date =		 2014,
  publisher =	 {{MIT Press}},
  url =			 {http://site.ebrary.com/id/10956108},
  urldate =		 {2020-04-08},
  isbn =		 {978-0-262-32052-8},
  langid =		 {english},
}

@incollection{lawvereCategoriesSpaceQuantity1992,
  title =		 {Categories of {{Space}} and of {{Quantity}}},
  booktitle =	 {The {{Space}} of {{Mathematics}}},
  author =		 {Lawvere, F. William},
  editor =		 {Echeverria, Javier and Ibarra, Andoni and Mormann, Thomas},
  date =		 {1992-01-31},
  publisher =	 {{de Gruyter}},
  location =	 {{Berlin, Boston}},
  doi =			 {10.1515/9783110870299.14},
  url =			 {https://www.degruyter.com/view/books/9783110870299/9783110870299.14/9783110870299.14.xml},
  urldate =		 {2020-04-13},
  isbn =		 {978-3-11-087029-9},
  langid =		 {english},
}

@book{corfieldModalHomotopyType2020,
  title =		 {Modal {{Homotopy Type Theory}}: {{The Prospect}} of a {{New Logic}} for {{Philosophy}}},
  shorttitle =	 {Modal {{Homotopy Type Theory}}},
  author =		 {Corfield, David},
  date =		 {2020-02-06},
  publisher =	 {{Oxford University Press}},
  location =	 {{Oxford, New York}},
  isbn =		 {978-0-19-885340-4},
  pagetotal =	 192
}

@article{blackHoleModelOfComputation,
  title =		 "Black hole as a model of computation",
  journal =		 "Results in Physics",
  volume =		 13,
  pages =		 {{102--188}},
  year =		 2019,
  issn =		 "2211-3797",
  doi =			 {10.1016/j.rinp.2019.102188},
  url =			 {http://www.sciencedirect.com/science/article/pii/S2211379719304036},
  author =		 "G.R. Andrews"
}

@article{Schwarzschild:1916ae,
  author =		 "Schwarzschild, Karl",
  title =		 "{On the gravitational field of a sphere of incompressible fluid according to Einstein's
                  theory}",
  eprint =		 "physics/9912033",
  archivePrefix ="arXiv",
  journal =		 "Sitzungsber. Preuss. Akad. Wiss. Berlin (Math. Phys.)",
  volume =		 1916,
  pages =		 "424--434",
  year =		 1916
}

@article{Schwarzschild:1916uq,
  author =		 "Schwarzschild, Karl",
  title =		 "{On the gravitational field of a mass point according to Einstein's theory}",
  eprint =		 "physics/9905030",
  archivePrefix ="arXiv",
  journal =		 "Sitzungsber. Preuss. Akad. Wiss. Berlin (Math. Phys.)",
  volume =		 1916,
  pages =		 "189--196",
  year =		 1916
}

@article{rosalesGeneralTheoreticalFramework2019,
  title =		 {A General Theoretical Framework for the Design of Artificial Emotion Systems in {{Autonomous
                  Agents}}},
  author =		 {Rosales, Jonathan-Hernando and Rodríguez, Luis-Felipe and Ramos, Félix},
  date =		 {2019-12-01},
  journaltitle = {Cognitive Systems Research},
  shortjournal = {Cognitive Systems Research},
  volume =		 58,
  pages =		 {324--341},
  issn =		 {1389-0417},
  doi =			 {10/ggh7sr},
  url =			 {http://www.sciencedirect.com/science/article/pii/S1389041719304619},
  urldate =		 {2020-01-21},
  langid =		 {english}
}
\end{document}